\documentstyle[12pt,aasms4] {article}

\lefthead{R. Schild}
\righthead{Q0957+561 Brightness Record Wavelet Analysis}

\begin{document}

\title{A Wavelet Exploration of the Q0957+561 A,B Brightness Record}

\author{R. Schild}

\affil{	Center for Astrophysics, 60 Garden Street, Cambridge MA 02138\\
		rschild@cfa.harvard.edu}
 
\begin{abstract}

	We analyze the Q0957+561 A,B brightness record with 3 types of
	wavelets to define properties that are
	independent of the nature of the analyzing wavelet. The wavelet
	analysis picks out features having arbitrary shape, and localizes
	them in time and fits amplitudes.

	We find that independently of the analyzing wavelet, the mean
	wavelet amplitude is zero, meaning that there are as many positive
	as negative brightness spikes. For all wavelet durations the fitted
	amplitudes are equal in the A and B images, except that image B may
	have larger amplitude brightness fluctuations for the longest
	duration 64-day features. Independently of the wavelet family
	fitted or of the kind of statistical measure of wavelet amplitude,
	the fitted amplitudes seem to increase as a linear function of the
	wavelet duration, with the mean absolute deviation about a factor
	of 10 greater for 64-day wavelets than for 2-day wavelets. The
	underlying physical process producing the brightness fluctuations
	is found to have more power in long duration wavelets than in short
	duration wavelets than a process dominated by white noise. Thus it
	is established that the rapid brightness fluctuations observed in
	the Q0957 images A and B are not the result of observational noise.

\end{abstract}


\section{Introduction: Why Wavelet Analysis?}

	The Q0957+561 brightness record provides a unique data stream for
the exploration of quasar brightness fluctuations modified by
microlensing. The data have been consistently reported from CCD
observations since the object's discovery in 1979, and by now data for 1200
nights of observation are available, with a nightly average brightness
reported by Schild and Thomson (1996) and earlier references. 
Although the original 1986 Schild and Cholfin time delay has long been
controversial, it is now confirmed by radio and optical data (Haarsma et al
1997).  The optical CCD data have recently
been analyzed by Pelt et al (1996, 1998), Pijpers and Wanders (1997), 
Thomson and Schild (1997),
and Goicoechea et al (1998) for time delay between arrival of the two
quasar images. With the time delay now satisfactorily determined and
the cosmological implications for the determination of the Hubble Constant
amply discussed by Kundic et al (1997), it is time to consider other
results mined from the Q0957 brightness record.

	The most obvious feature of the Q0957 data record is the unexpected
rapid brightness fluctuations of low amplitude found. It has long been
known that these features cannot be intrinsic to the quasar, because if
they were the time delay would have been easy to determine. Many authors
have published cross-correlation or equivalent calculations for the
determination of time delay, and the Full Width at Half Maximum or some
equivalent measure of the width of the autocorrelation peak always is found
to be of order 100 days, as was found even in the first successful
autocorrelation calculation of Schild and Cholfin (1986). 
This probably means that the high frequency part of
the brightness record does not originate in intrinsic quasar brightness
fluctuations. A more direct demonstration of this is found in the
calculations of Pelt et al (1996, Figs. 10 and 11),
where the brightness fluctuation time series was
filtered and separated into high-frequency and low-frequency components.
These authors then demonstrated that the high-frequency part of the
brightness record does not show the ~415 day time delay; the low frequency
part does.

	Because the high frequency component of the quasar's brightness
record does not display the time delay, it seems likely it originates  	
in microlensing, although the exact mechanism is unclear. Schild(1996) has
discussed the probable origin in a population of planetary mass objects
in the lens galaxy, but has noted that for such an explanation to be viable
the quasar source must have structure on a scale comparable to the size of
the Einstein ring for a $10^{-6}M{\sun}$ microlens in the lens galaxy. Gould and
Miralda-Escude (1997) concur with the general findings of Schild regarding the
microlensing origin, and suggest that some quasar source structure may be
in motion to modify the relevant time scales. Because no detailed model
calculation of the rapid microlensing exists, it would have to be said that
the nature of the microlensing is as yet not understood. Schild (1996)
considers the possibility that the fluctuations originate in part in mass
particles in the Halo of our Galaxy, or in the quasar host galaxy, 
and a microlensing calculation with two lens screens may be
required. Given that Schild (1996) and Gould and Miralde-Escudo (1997) 
all attribute the rapid brightness 
fluctuations to microlensing, for the remainder of this report
we refer to them as microlensing.

	Although questions remain about the accuracy of the Mount Hopkins
data which define the microlensing, in the one situation where direct
comparison with other observatories is possible, the data and the rapid
microlensing are confirmed (Gibson and Schild 1999). A more comprehensive
discussion of the microlensing has been given by Schild (1996), who finds
that the microlensing is seen as a
continuous pattern of brightness peaks having  cusp-shaped profiles and
amplitudes of several percent (see Schild 1996 Fig. 2). 
An excellent example of a
microlensing event seen in the brightness record of the A image is given as
Fig. 5 of Schild 1996. That profile shows a 90-day event well above the
observational noise, together with a nearly continuous pattern of lower
amplitude events which are barely recognizable above the noise of the
brightness detection. These weak features are admittedly poorly determined,
because their amplitude is barely, or not even, above the noise level,
shown as the grey zone in that illustration. However, even though an
individual data point is in the grey noise zone, the data points form
distinctive trends that indicate cusp-shaped profiles in a nearly
continuous pattern. The situation is similar to Q2237, where it is
generally considered that microlensing has been observed because even
though the individual data points do not significantly prove microlensing,
their overall trend is considered convincing (Corrigan 1991). 
And in the one case where
the quasar has been amply observed by 3 observatories concurrently, a
30-day microlensing event was distinctly seen (Gibson and Schild 1999).

	An inspection of the microlensing record of Fig. 2 in Schild (1996)
seems
to suggest that the strongest microlensing events have a peak amplitude of
about 0.10 mag and duration of about 1/4 year. Their profiles seem to be
more cusp-like than sinusoidal. Thus it would seem {\it ab initio} 
that they are
better described by wavelets than by sines and cosines. Moreover, in
the present writing we are not concerned about periodic effects in the cusp
features, but rather just in the most elementary statistical descrption of the
amount of power as a function of duration of the features. Because a wavelet
analysis provides the mathematical tools to perform a multiscale
correlation analysis with an analyzing function similar to the details of
the observed and expected fluctuations, it provides an optimum methodology
for our application.
 
	The first wavelet analysis applied to the Q0957 system was reported
by Hjorth et al(1992) for the determination of time delay. 
The conclusion of the
application was that the time delay was not determined, but this report
contains an excellent statement of the nature of wavelet analysis. This
author also finds particularly useful the book Wavelets in Geophysics by
Foufoula-Georgiou and Kumar (1994) for examples of applications in physical
science, with many references to more theoretical texts. Early developments
of the wavelet theory were given by Mallat(1989), and the first
application to problems in astronomy was by
Slezak, Bijaoui, and Mars (1990); the first review
of wavelet analysis in astronomy was given by Scargle (1997),

	Although our discussion is oriented toward microlensing, we in fact
analyze the separate A and B brightness records. While in principal
microlensing should be studied by subtracting the time delay shifted B from
the A component brightness record, to do so loses information about
which image causes the fluctuations. We have previously noted that
the quasar's rapid brightness fluctuations on time scales up to 100 days 
seem to be dominated by microlensing and our purpose here is to
demosnstrate properties of rapid brightness fluctuations in the two image
components. 

\section{The Data Set}

	Because our basic data sampling is once nightly, a Nyquist
argument gives our basic wavelet resolution lower limit of 2 days, and this
is the shortest duration wavelet we will fit to the data. Optimum fitting
dictates that wavelets of increasing powers of two be subsequently fitted,
so the wavelets are of duration 2, 4, 8, 16, 32, ... days. A major problem
is that the data record has many missing nights, and to minimize this
problem we investigate only the data subset of the last four observing
seasons, when observations were attempted nightly during the 9-month
observing season. As will be discussed below, corrections for missing data
will be derived and applied to the results. The observing season for Q0957
starts on Oct 1 and ends June 30 the following year. Data for the 1992-93
and 1993-94 have been published by Schild and Thomson (1996) and data for
the two following seasons are available in tabular from from the author's
homepage http://cfa-www.harvard.edu/rschild/ 
and have already been analyzed by some authors (Goicoechea et al 1998).

	Because the data are incomplete they will need to be interpolated
for the present program. We have of course experimented with several
methods of interpolation over the years, but simple linear interpolation
seems best for the present effort. Linear interpolation cannot increase
wavelet power at any frequency, whereas other schemes might. And with
linear interpolation, corrections for missing data are conceptually and
computationally simplest.
 
\section{Analysis}

	As yet no particular wavelet or wavelet family has been shown to be
optimal for microlensing investigations, and we will try three examples
known to have different properties to see if any significant differences
are found. The wavelet literature is full of considerations of
orthonormality of wavelets, which relates to optimum compression of signals
or optimum signal reconstruction. We are more simply concerned with how
effectively the various wavelet types fit our cusp-shaped data profiles,
and our approach will be to try three different choices to see whether and
how the results agree. However we are also anticipating future developments
and will concentrate on wavelets with the most useful properties

	Our first choice will be the Coiflet 2 (coif2) wavelet displayed
in Fig 1a. It is a nearly symmetrical wavelet with a scaling function,
also shown in Fig 1a.
As with all wavelets, it has positive and negative components so its
integral is unity; of course the positive spike predominates in the fitting
process. The Coiflet wavelets are orthogonal, biorthogonal, and have
compact support; they are useful for discrete wavelet transforms and 
for continuous wavelet transforms. They are generally regarded as compactly
supported wavelets with the highest number of vanishing moments for both
$\phi$  and $\psi$ for a given support width.

	Our second two choices are asymmetrical and part of the orthonormal
set of wavelets by Daubechies. As illustrated , the Daubechies 2
(Db2 Fig 1b) and Daubechies 3 (Db3, Fig 1c)
wavelets are asymmetrical and have negative side-lobes of different
character. Our purpose in including these two alternatives is to test
whether the wavelet fits give results which are significantly dependent on
the wavelet type. Like the Coiflets, the Daubechies wavelets are known as
compactly supported wavelets with the largest number of vanishing moments
for a given support width. They are orthogonal, biorthogonal, have compact
support, and are useful for discrete and continuous wavelet transforms.
Their associated scaling filters are minimum-phase filters.

	Figure 2 shows the wavelet decomposition of the Q0957+561 A
image for a coif2 wavelet, and in Fig. 3 shows the decomposition for
image B. In both figures, the signal analyzed {\it s} is shown in the upper
left panel and is the brightness record for the most recent 4 years when
the quasar observation was scheduled nightly. 
Data from multiple images obtained the same night were averaged 
together for a single nightly average. In all of the panels of Figs. 2-4,
the amplitudes of the wavelet coefficients and the residuals, as well as
the original signal, are expressed in magnitude units. The zero
point for the dates of observation was JD 2448901. 
Figures 2 and 3 show the decomposition of the A and B brightness records into
wavelets of duration 2-64 days, plus a residual signal a6, as explained in
the Figure 2 caption. The power of this approach is that the six detail
signals d1...d6
now provide the basis for detailed estimation of numerous properties of the
original signal, such as the power in negative {\it vs} positive wavelets
and the amount of wavelet power as a function of wavelet duration. Such
properties obviously relate to the physical process producing the
brightness fluctuations, and our purpose is to provide a quantitative
means of comparing observations to model simulations.

	Before deriving a catalogue of wavelet properties and determining
and applying corrections for the incompleteness of the data set in
subsequent sections, we will provide an additional illustration to
display some interesting properties of the wavelet fits. The impatient or
sophisticated reader may prefer to skip to the next section.  

	Figure 4 shows the lowest order wavelet decomposition of a
subset of the A data record analyzed previously. The date zero point and
scale are the same as in Fig 2, and the data and results may be recognized
as basically a zoom of a portion of that illustration. The subset was
chosen because of the extraordinary quiescence of the quasar to daily
brightness fluctuations, even though fluctuations on a time scale of 4
days, shown in the d2 box, were normal. From this data subset we conclude
that the Mt Hopkins data are capable of recording very quiescent data at
some frequencies, even as the fluctuations were more normal at lower
frequencies. This has implications about the accuracy of the data, since to
record quiescence at high frequencies requires both that the source be
quiescent and that the data be of high quality. Schild (1990)
had previously noted the surprising quiescence of the A image in 1988-89, and
we may now consider this quasar/microlensing property 
confirmed from observations in the 1995-96 season. 
As regards accuracy of the
data, we have long considered that the largest uncertainty in our
monitoring data is from correlated power between the two closely spaced
(6 arcsec separation) images due to seeing effects, as discussed in
Appendix 1 of Schild and Cholfin (1986). However this effect should have
its characteristic signature; if seeing gets worse, the enlarged quasar
images spill over into one another and both images are measured to be too
bright (recall that the basic photometric procedure is to measure the
total luminosity contained within a 6 arcsec diameter aperture with the
aperture photometry of 5 standard stars lacking close companions). Thus
deteriorated seeing always produces comparable photometric errors in both
images, and always in the sense of excess brightness. Because poor seeing
spells at Mt Hopkins are usually experienced on time scales of a single
night or portion of a night, correlated errors due to seeing effects should
be seen on single day time scales. We may conclude from Fig. 4 that 
since only very low amplitude fluctuations are found most of the time, seeing
effects evidently do not affect the Mt Hopkins data at the 0.005 mag level.
Comparison of the A and B brightness records for the period, as shown in
Figs 2 and 3, shows that the B component showed the usual brightness
fluctuations during the period.

A possibly troubling point also evident from inspection of figures 2 and 3
is the appearance that linear interpolation has caused large wavelet power
at the ends of the interpolation intervals. These features are seen around
date 1140 in fig. 2 and 1160 in Fig. 3. But a closer inspection will show
that these features are well embedded in real data, as confirmed by our
expanded scale calculations (not shown here). We in fact find no problem at
the ends of the interpolation intervals.
   
\section{Results}

\subsection{The Mean Values}

	We are now in a position to use the wavelet decompositions of the
two time series of brightness data to ask simple questions about the
physical process. Inspection of the details coefficients d1...d6 
in Figs. 2-4 suggests
that about as many positive as negative wavelets have been fitted to the
brightness record. In other words, there are about as many downward as
upward brightness spikes. Is this apparent in the statistics?
	
	Table 1 shows the mean values of the wavelet decompositions
d1 - d6 for the three types of wavelets, and for each quasar image.
For each level of decomposition d1...d6
we calculate the mean value of fitted wavelets,
and the rms deviation of the mean. The quantity for which we compute the mean
value is the the amplitude of wavelets fitted to the time series, and
if there is any significant tendency for the physical process to produce
more positive than negative wavelets for some wavelet scale,
then the means for the decompositions
d1-d6 should differ significantly from zero independent of the shape of the
analyzing wavelet. 
We find that the tabulated rms deviations of the means are in
general larger than the means so we conclude no significant departures of
the means are found for any level of decomposition. A possible
exception is that for the finest details at level d1 (2-day wavelets) a
small trend toward negative means may be found, but since it is a
low-significance conclusion and is not seen in any of the other
decomposition levels d2-d6, we dismiss it for now as a probable fluke
(since in 12 statistical tests you would expect one 3-sigma event). The
visual impressions of the decompositions in Figs. 2 and 3 are that positive
and negative brightness spikes are about equally large and numerous for all
wavelet scales. 

Equal positive and negative brightness spikes are of fundamental importance
because most astrophysical processes at low surface optical depth generate
asymmetrical brightness curves. We know of two processes causing equal
positive and negative features: microlensing at surface optical depths near
unity (Schneider, Ehlers, and Falco, 1991, p. 343.), and microlensing by a
double screen of microlenses, so that a second screen can decrease or
increase the magnification caused by the first screen. We accept that it
remains to be conclusively demonstrated that the brightness fluctuations
seen in the two quasar images are dominated by microlensing, but we have
already presented arguments that they probably are.
 
	We do find in Table 1 a small tendency for the image B
means to be more
negative than positive, but in all cases the mean value is smaller than its
rms deviation, and we do not claim a significant result. Note that since
the wavelets are being fitted to data in magnitudes, a small negative mean
would indicate a small preponderance of spikes of brightness increase.

\subsection{The Mean Absolute Deviation}

	Table 2 shows the mean absolute deviation of the wavelet
coefficients as a
function of wavelet duration for the three wavelet types. The table lists
under the column labeling the detail level corresponding to Figs. 2-4 and
the wavelet duration expressed in days, the observed wavelet coefficients
expressed in magnitudes. Thus the upper left table entry .002284 means
that the mean absolute amplitude of Coif2 wavelets of 2-day duration fitted
to the data record is .002284 mag. Proceeding down an individual column of
Table 2, we list the mean absolute deviations for
the three wavelet types, then the mean value and the rms deviation of
the three values contributing to the mean. Then we list the derived
corrections for incomplete sampling which depend on the wavelet duration,
and finally we list the value of the mean absolute deviation corrected for
incomplete sampling.

	Figure 5 shows the means as a function of wavelet duration.
From these results we reach the important conclusion, seen to be
independent of wavelet type and seen to apply to both image components;
{\it the short-term quasar fluctuations increase in amplitude with
increasing wavelet duration.} We shall see below that this conclusion is
not modified by corrections for incompleteness of the data record. We also
find no significant difference in microlensing power between the A and B
quasar images, except possibly for the longest, 64-day, wavelets considered
here. 

\subsection{Corrections to the Mean Absolute Deviation for Incompleteness}

	The data in Figure 5 need correction for the fact that 
brightness monitoring was not possible during the summer months when the
source is too close to the sun, and during bad weather spells or when the
CCD camera was unavailable for one reason or another. The corrections will
depend upon the duration of the fitted wavelet. For example, for the 64-day
wavelets, data missing for days or weeks should not affect the fitted
profiles, and we expect only a correction for the long summer break. Thus
the brightness record should be about 3/4 complete, and we expect a
correction factor of approximately 4/3. For the shortest, 2-day wavelets
fitted, the correction factor is the reciprocal of the fractional number of
dates of observation, since a missing day's data means that the program
interpolates across the two adjacent data points, and cannot determine the
correct wavelet amplitude. Thus for 2-day wavelets, the correction is the
reciprocal fraction of the number of observation dates in the observing
record, which is 471/1358. 
For wavelets of intermediate duration, the
missing data corrections are similarly computed from the number of missing
half-duration intervals, and the corrections are listed in Table 2 in the
rows marked {\it corr}.

	The Mean Absolute Deviation results for the wavelet fits are listed in
Table 2 in the rows marked {\it crtd} and are plotted in Figure 6, where 
the amplitudes are shown as ordinates for abscissas plotted as the wavelet
duration.  The error bars express rms deviation of a mean value for the
three wavelet types, to give some sense of the dependence of the mean
absolute deviation on wavelet choice.
In this plot we find a striking result that {\it the amplitudes are
simply proportional to wavelet duration}. There is some suggestion that for
image B, an additional component may be present at the longest, 64-day
wavelets.  If T is the wavelet duration in days and W is the Mean Absolute
Deviation listed in the {\it ctd} rows of Table 2, then W = .00101(T) +
.0063. As noted above, image B may have a slightly higher value for
wavelets 64 days and longer.

	The characterization of the underlying process producing brightness
fluctuations in Fig. 6 is easily seen to differ from white noise, for 
which the average wavelet coefficients would be equal for all wavelet
durations. In the Q0957 data set observational errors were estimated based
upon multiple observations each night. If these observational errors had
been significantly underestimated, to the extent that the rapid brightness
fluctuations reported are dominated by Gaussian statistical errors, the
resulting process would be described as white noise. For a white noise
process, the mean wavelet amplitudes expressed as either a mean absolute
deviation or as a root mean square would be equal at all wavelet scales,
which means that the line connecting the observational means in figures 5 -
8 would be a horizontal straight line. Such a line cannot fit the data, and
we conclude that observational errors do not dominate the observed pattern
of brightness fluctuations on time scales investigated here.

\subsection{The rms Mean}

	As a means of testing the robustness of our approach and
conclusions, we consider a second measure of amplitude of the fitted
wavelets. In the previous discussion, a linear measure of amplitude of the
fitted wavelengths was adopted with the mean absolute deviation; we now
adopt a quadratic measure, the root mean square (rms) deviation. This is
the familiar statistical estimator with the well known properties that all
deviations from the mean are squared and therefore positive, and that the
several strongest peaks contribute most to the rms measure. Our treatment
proceeds analogously to that for the MAD measure, and the new analogues to
Table 2 and Figs 5 and 6 are Table 3 and Figs 7 and 8.

	Figure 7  shows the Table 3 data for the uncorrected wavelet
amplitudes (not corrected for incomplete sampling). Compared to figure 5, we
find a larger scatter for the amplitudes expressed as an rms, especially at
32 and 64 day wavelets, presumably because the rms averages are more
sensitive to the several largest amplitude wavelets. This sensitivity
probably also accounts for the large discrepancy at 64-days between the
amplitudes for the different wavelet types on the B image
data. Nevertheless as was found for the MAD amplitude measure, the B quasar
image seems to have more wavelet amplitude than image A for the
longest-duration wavelets. 

	In correcting the rms data to be plotted as Fig. 8 a procedural
change must be made appropriate to the rms statistic. The corrections for
the number of missing data points must now be entered with a harmonic mean
instead of a linear multiplier. With this modification the {\it crtd}
data in Table 3 and Fig. 8 are obtained.

	The corrected rms wavelet amplitudes shown in Fig 8 show one
potentially important feature found previously in Fig. 6; the amplitudes of
the wavelets agree to within the errors for all but the longest wavelets,
and for these 64 day wavelets the B quasar image has the greatest
amplitude. Recall that the B image is seen closer to the lens galaxy G1's
image, so image B has an optical depth to microlensing a factor approximately
3.3 larger than image A. 

	A property also common to both measures of wavelet amplitude is the
upturn from a linear fit that occurs for the shortest, 2-day wavelets.
This feature suggests that some excess power at the shortest 1-day
timescale in the data set has a significant contribution from observational
errors. This property was discussed in 3, and is somewhat dependent on the
accuracy of the corrections applied for data incompleteness. We expect to
return to this point in a future report in which we will examine several
80-day data subsets which have more complete sampling on the daily time scale.

	From a fit to our corrected Table 3 data, as shown in Fig. 8, we
determine a best fitting mean relation W = .00133 * T + .0060, with
variables as defined in Section 4.3.
We find for the wavelet coefficients expressed as a root mean square as we
found for mean absolute deviation that the underlying process differs from
a white noise process, which produces equal rms coefficient amplitudes at
all wavelet scales.

\section{ Summary and Conclusions}

	In our first wavelet decomposition of the
Q0957 brightness record, we have adopted the viewpoint that the predominant
fluctuations originate in microlensing, not in intrinsic quasar
fluctuations. This follows as a conclusion from the Pelt et al (1995) 
attempt to find the time delay in the filtered, high frequency portion of
the brightness record, and the common experience that the width of any
cross-correlation plot yet published is of order 100 days. In future
reports we expect to test the assumption further by autocorrelating the
wavelet fits to the A and B quasar images to define the fractional quasar
{\it vs} microlensing amplitude at all time scales.

	In order to make wavelet decompositions at all time scales we have
linearly interpolated the brightness record and then applied corrections
for incomplete sampling. In a future report we expect to check these
corrections by investigating subsamples of the data with nearly complete
sampling. We will also expand our analysis to longer duration wavelets, by
analyzing the total 18-year data record rather than the 4-year subsample
studied here. We believe that our wavelet fits and missing data corrections
are correct to within a factor of two, and this will already allow
significant comparisons to microlensing models (which do not yet exist for
this system).

	Subject to these uncertainties, we report the following
conclusions: 

1. The mean wavelet amplitude at all time scales is zero, meaning that
there are as many negative as positive wavelets.

2. The fitted wavelet amplitudes are equal in the A and B images, except
that for the longest fitted wavelets of 64 day duration the B image may
have larger amplitude brightness fluctuations.

3. Independently of the family of wavelet fitted or the statistical measure of
wavelet amplitude, the amplitudes seems to increase as a linear function
of wavelet duration, with mean absolute deviation a factor of approximately
10 greater for 64-day wavelets than for 2-day ones. A white noise process
causing the brightness fluctuations would produce equal mean coefficient
amplitudes at all scales, and we conclude that the rapid brightness
fluctuations observed in the A and B quasar images are not dominated by
observational noise.

\section {Acknowledgements}

I wish to thank Dr Eric Kolaczyk for helpful remarks about applications of
wavelets. 

\vfill\eject

\clearpage
\centerline{\bf FIGURE CAPTIONS}

Figure 1a. Wavelet properties for the Coiflet2 (coif2) wavelet. The wavelet
shape, called the wavelet function $\psi$, is shown in the upper right panel,
and the upper left panel shows the associated scaling function $\phi$. 

Figure 1b. Wavelet properties for the Daubechies 2 (Db2) wavelet, as
described in Fig 1a.

Figure 1c. Wavelet properties for the Daubechies 3 (Db3) wavelet, as 
described in Fig 1a.

Figure 2. A full wavelet decomposition for the 4-year data subset of the
A (northern) quasar image with a second order Coiflet (Coif2) analyzing
wavelet. The signal analyzed in the upper left panel is labeled {\it s}.
The lower right
panel labeled d1 shows the details of fitting the wavelets with a 2-day
width to the signal, and the lower left panel labeled a1 shows an
approximation to the signal with the details of d1 removed. So adding
coif 2 wavelets of amplitude given in d1 to the residual signal a1 produces
an exact reconstruction of the signal s. Similarly panel d2 shows the
details of a wavelet fit to a1, and the residual signal with the d2
wavelets removed is seen in panel a2. An exact reconstruction of signal
s is obtained by adding to signal a2 a pattern of coif 2 wavelets having 4 day
duration and amplitudes d2, together with a pattern of 2-day duration
wavelets having amplitudes given in d1. Similarly, panels d3...d6 show the
patterns of wavelets which when fitted to the signal give approximations
a3...a6. The original signal is thus decomposed into the 6 signals d1...d6
by which coif2 wavelets of duration 2...64 days are multiplied and added
to residual signal a6 for an optimum wavelet representation of the original
signal. The power of this approach is that the six detail signals d1...d6
now provide the basis for detailed estimation of numerous properties of the
original signal, such as the power in negative {\it vs} positive wavelets
and the amount of wavelet power as a function of wavelet duration. Such
properties obviously relate to the physical process producing the
brightness fluctuations, and our purpose is to provide a quantitative
means of comparing observations to model simulations.

Figure 3. A Coif2 full wavelet decomposition of the 4-year data subset of the
B (southern) quasar image.

Figure 4. A wavelet analysis of a quiet subset of the A brightness record.
The bottom panels d1 and d2 show the Coif2 wavelets of 2- and 4-day width
fitted to the A quasar image signal {\it s}. 

Figure 5. The mean absolute deviation measure of wavelet amplitude as a
function of wavelet duration for the A and B quasar images, and for the
three analyzing wavelets as identified. Thus the curve marked B-Db3 shows
the wavelet amplitude for the B quasar image with the Daubechies 3
analyzing wavelet.

Figure 6. The corrected mean absolute deviation measure of wavelet
amplitude for the Coif2 analyzing wavelet applied to the A and B quasar
images, with corrections for missing data applied as noted in the text.

Figure 7. The standard deviation measure of wavelet amplitude as a function
of wavelet duration, with the several curves identified as in Fig. 5

Figure 8. The corrected standard deviation measure of wavelet amplitude for
the Coif2 analyzing wavelet applied to the A and B quasar images, with
corrections for missing data and interpolation applied as noted in the text.

\clearpage

\begin{deluxetable}{lrrrrrrr}
\tablenum{1}
\tablewidth{0pt}
\tablecaption{Mean values}
\tablehead{
\colhead{ Image} & \colhead { Wavelet} &\colhead {d1} & \colhead{d2}
&\colhead{d3} &\colhead {d4} &\colhead {d5} &\colhead { d6 }}
\startdata A & Coif2 & -.00702 & -.02151 & -.00141 & .00018 & .00197 & .01187
\nl
\nodata & Db2 & -.01080 & .04512 & -.00186 &  .04895 & -.00869 & -.00011 \nl
\nodata & Db3 & -.00771 & -.00013 &  .00026 & .00052 & -.00128 & -.00978
\nl
\nodata & mean & -.00851 & .00783 & -.00100 & .01655 & -.00267 & .00066 \nl
\nodata & rms & .00201 & .03402 & .00112 & .02806 & .00546 & .01085 \nl
B & Coif2 & -.01238 & -.04371 & -.00139 & -.00369 & .00044 & .00437 \nl
\nodata & Db2 & -.02449 & .00037 & -.00148 & -.00347 & -.00209 & -.01903
\nl
\nodata & Db3 & -.00674 & -.00036 & -.03605 & .00028 & -.00050 & -.02353
\nl
\nodata & mean & -.01454 & -.01457 & -.01297 & -.00229 & -.00072 & -.01273
\nl 
\nodata & rms &.00907 & .02524 & .01999 & .00223 & .00128 & .01498 \nl
\enddata
\end{deluxetable}

\clearpage
\begin{deluxetable}{lrrrrrrr}

\tablenum{2}
\tablewidth{0pt}
\tablecaption {Mean absolute deviation}
\tablehead{ 
\colhead{Image} & \colhead{Wavelet} & \colhead{d1} & \colhead{d2} & 
\colhead{d3} & \colhead{d4} & \colhead{d5}  & \colhead{d6}} 

\startdata A & Coif2 &  .002284 & .005104 & .01046 & .01732 & .03142 &
.05854 \nl
\nodata & Db2 & .002492 & .005236 & .01086 & .01698 & .03585 & .05583 \nl
\nodata & Db3 & .002434 & .005345 & .00916 & .01811 & .03566 & .06099 \nl
\nodata & mean & .00240 & .00523  & .01016 & .01747 & .03331 & .05845 \nl
\nodata & std  & .00011 & .00012  & .00089 & .00058 & .00216 & .00258 \nl
\nodata & corr & 2.88   & 1.736   & 1.536  & 1.377  & 1.226  & 1.165  \nl
\nodata & crtd & .00691 & .00908  &.01561  & .02406 & .04058 & .06804 \nl
B       & Coif2 & .002126 & .004785 & .01059 & .01703 & .02495 & .06409
\nl
\nodata & Db2 & .002254 & .005669 & .01072 & .01671 & .03822 & .07733 \nl
\nodata & Db3 & .002258 & .005077 & .01001 & .01792 & .03057 & .09169 \nl
\nodata & mean & .00221 & .00518  & .01044 & .01722 & .03125 & .0776  \nl
\nodata & std & .00008  & .00045  & .00038 & .00063 & .00666 & .01380 \nl
\nodata & corr & 2.88   & 1.736   & 1.536  & 1.377  & 1.226  & 1.165  \nl
\nodata & crtd & .00636 & .00899  & .01604 & .02371 & .03831 & .09051 \nl
\enddata

\end{deluxetable}
\clearpage

\begin{deluxetable}{lrrrrrrr}
\tablenum{3}
\tablewidth{0pt}
\tablecaption {Std. deviation measure of wavelet amplitudes}
\tablehead{
\colhead{Image} & \colhead{Wavelet} & \colhead{ d1} & \colhead { d2} &
\colhead {d3} & \colhead{ d4} & \colhead{ d5} & \colhead {d6}}
\startdata
A & Coif2 & .005315 & .008599 & .01622  & .02378  & .04586  & .08241 \nl
\nodata & Db2 & .005421 & .009639 & .01722 & .02524 & .04837  & .07441 \nl
\nodata & Db3 & .005288 & .009154 & .01360 & .02706 & .04955  & .08056 \nl
\nodata & mean & .00534 & .00913  & .01567 & .02536 & .04793  & .07913 \nl
\nodata & std  & .00007 & .00052  & .00186 & .00164 & .00188  & .00419 \nl
\nodata & corr & 2.1294 & 1.2416  & 1.1346 & 1.0687 & 1.0252  & 1.0135 \nl
\nodata & crtd & .01137 & .01134  &.01778  & .02710 & .04914  & .08020 \nl
B      & Coif2 & .00479 & .00784  &.01604  & .02457 & .03566  & .09171 \nl
\nodata & Db2  & .00482 & .009510 & .01594 & .02235 & .05422  & .1033  \nl
\nodata & Db3  & .00453 & .008360 & .01512 & .02727 & .04060  &.12322  \nl
\nodata & mean & .00472 & .00857  & .01570 & .02473 & .04349  & .10607 \nl
\nodata & std  & .00016 & .00085  & .00050 & .00246 & .00961  & .01593 \nl
\nodata & corr & 2.1294 & 1.2416  & 1.1346 & 1.0687 & 1.0252  & 1.0135 \nl
\nodata & crtd & .01005 & .01064  &.01781  &.02643  & .04459  & .10750 \nl
\enddata
\end{deluxetable}

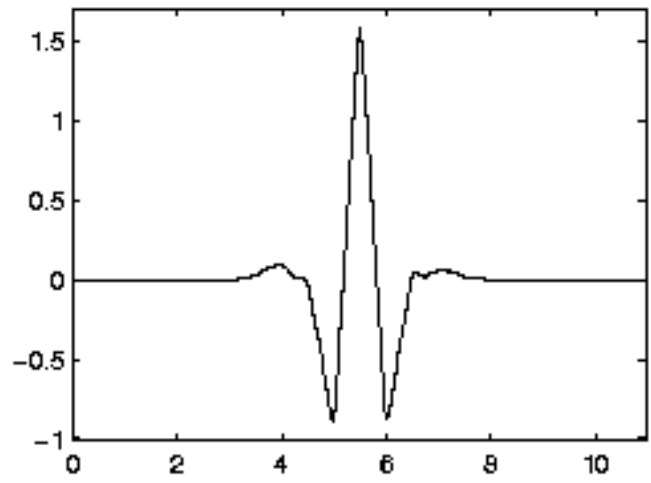
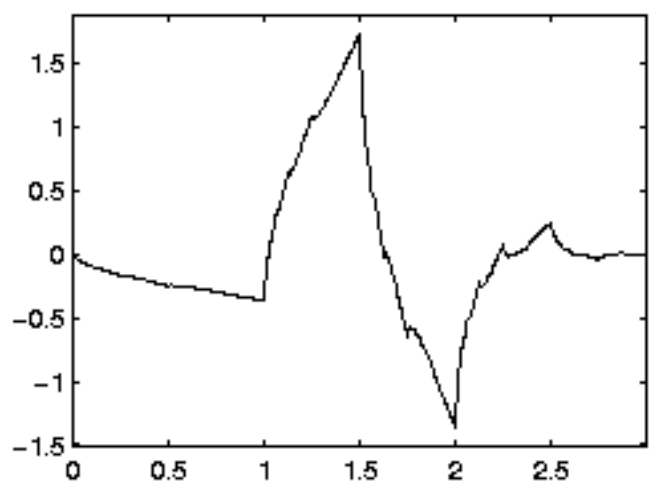
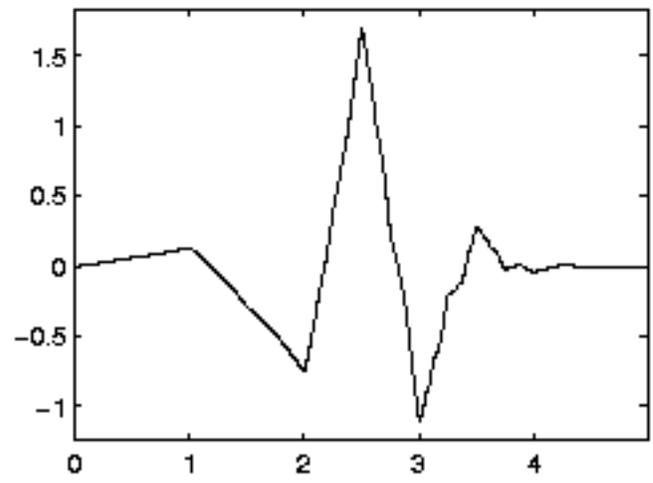
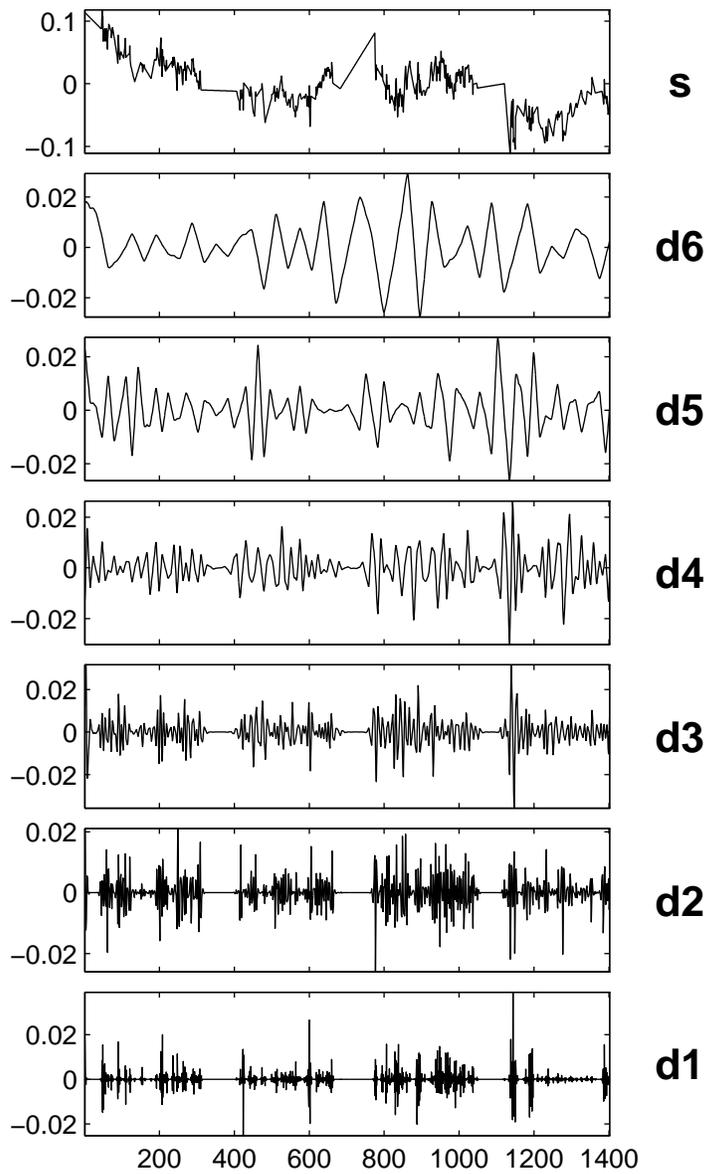
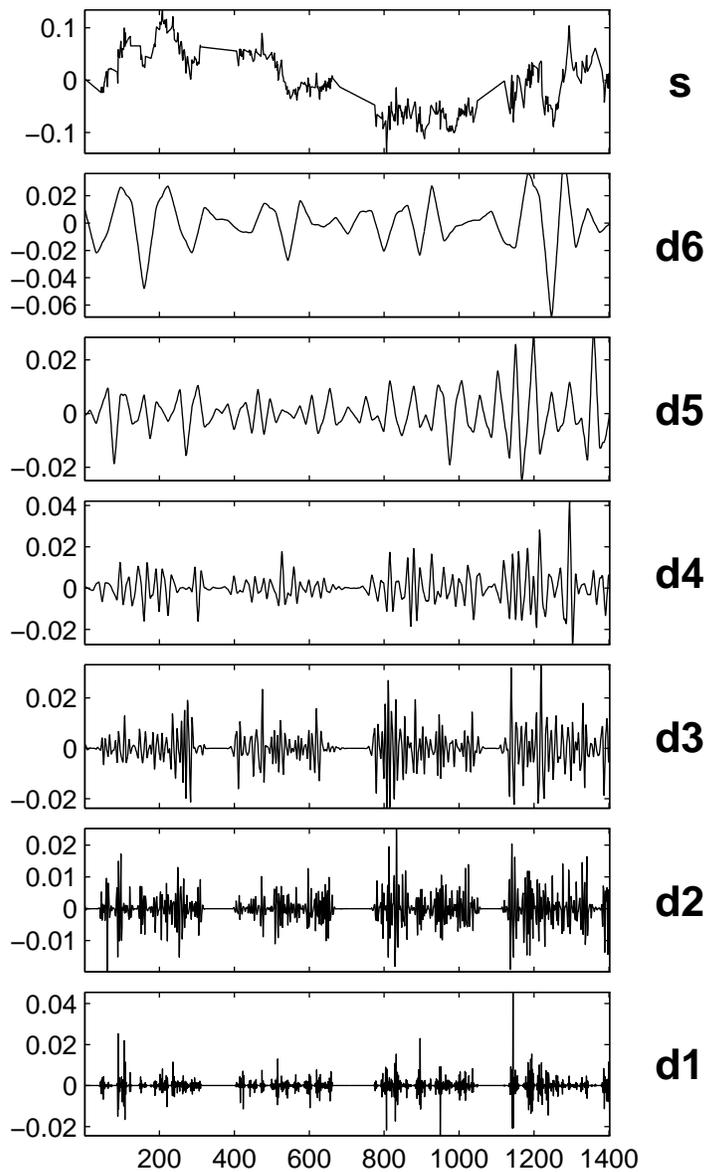
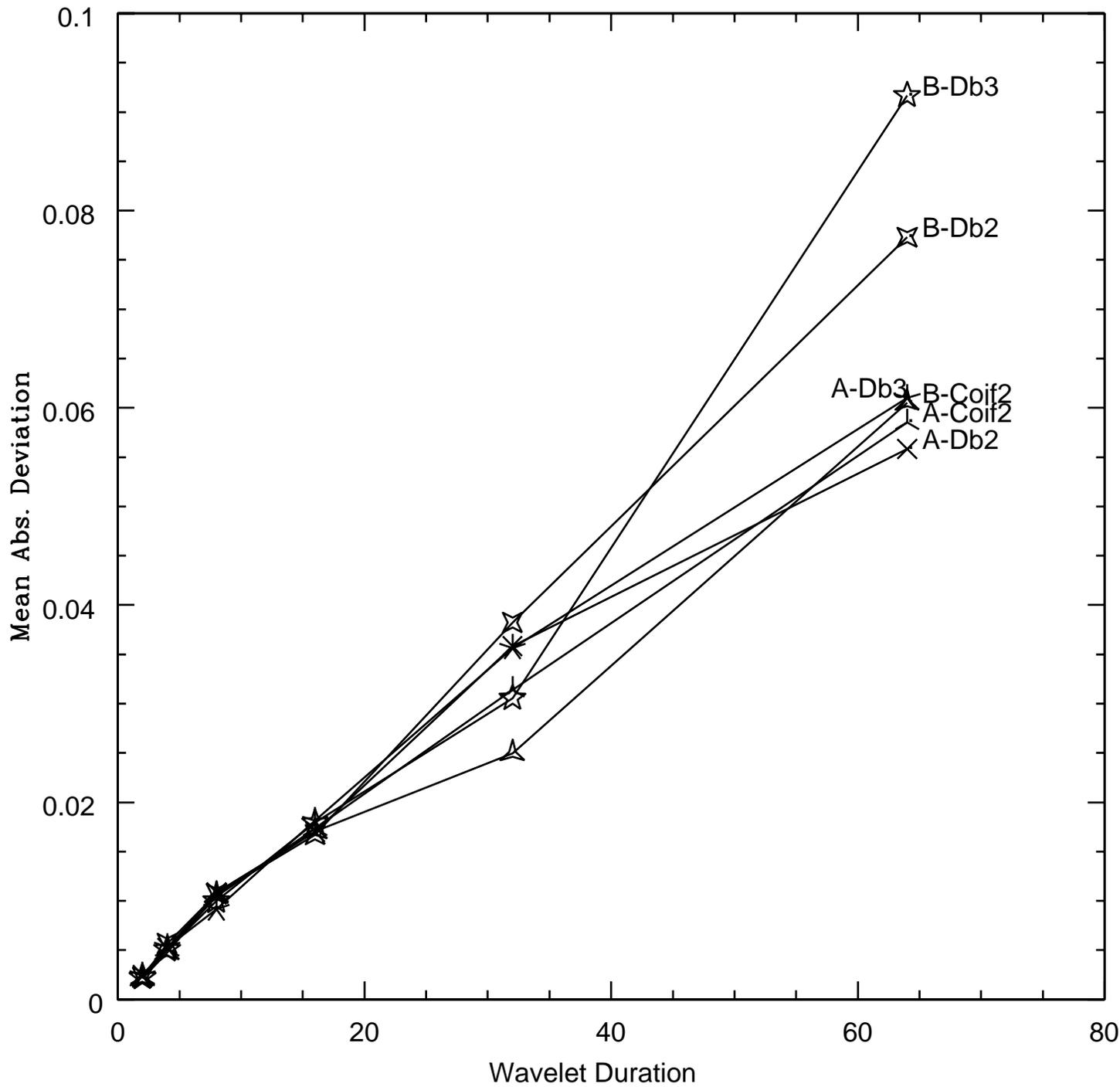
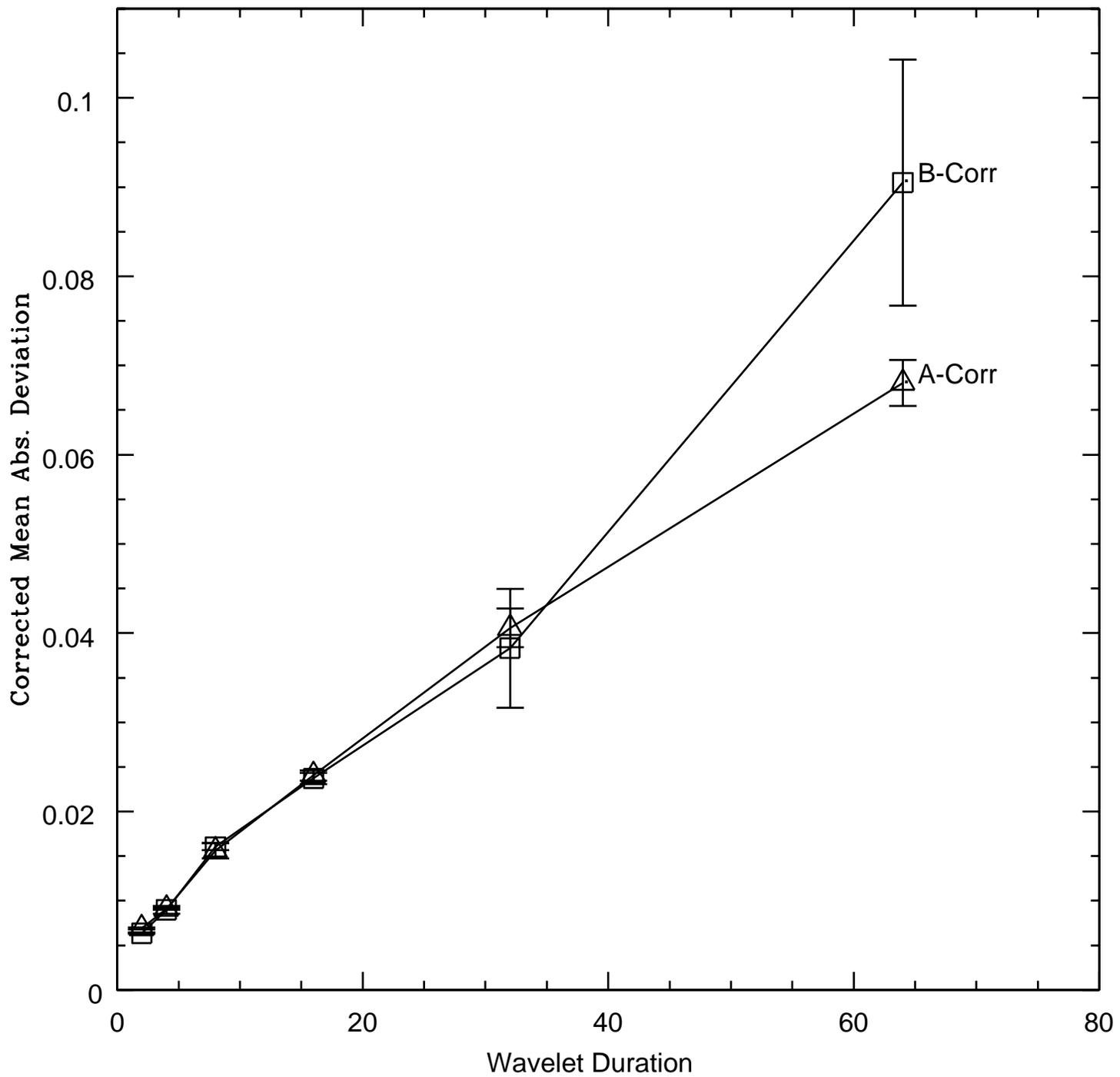
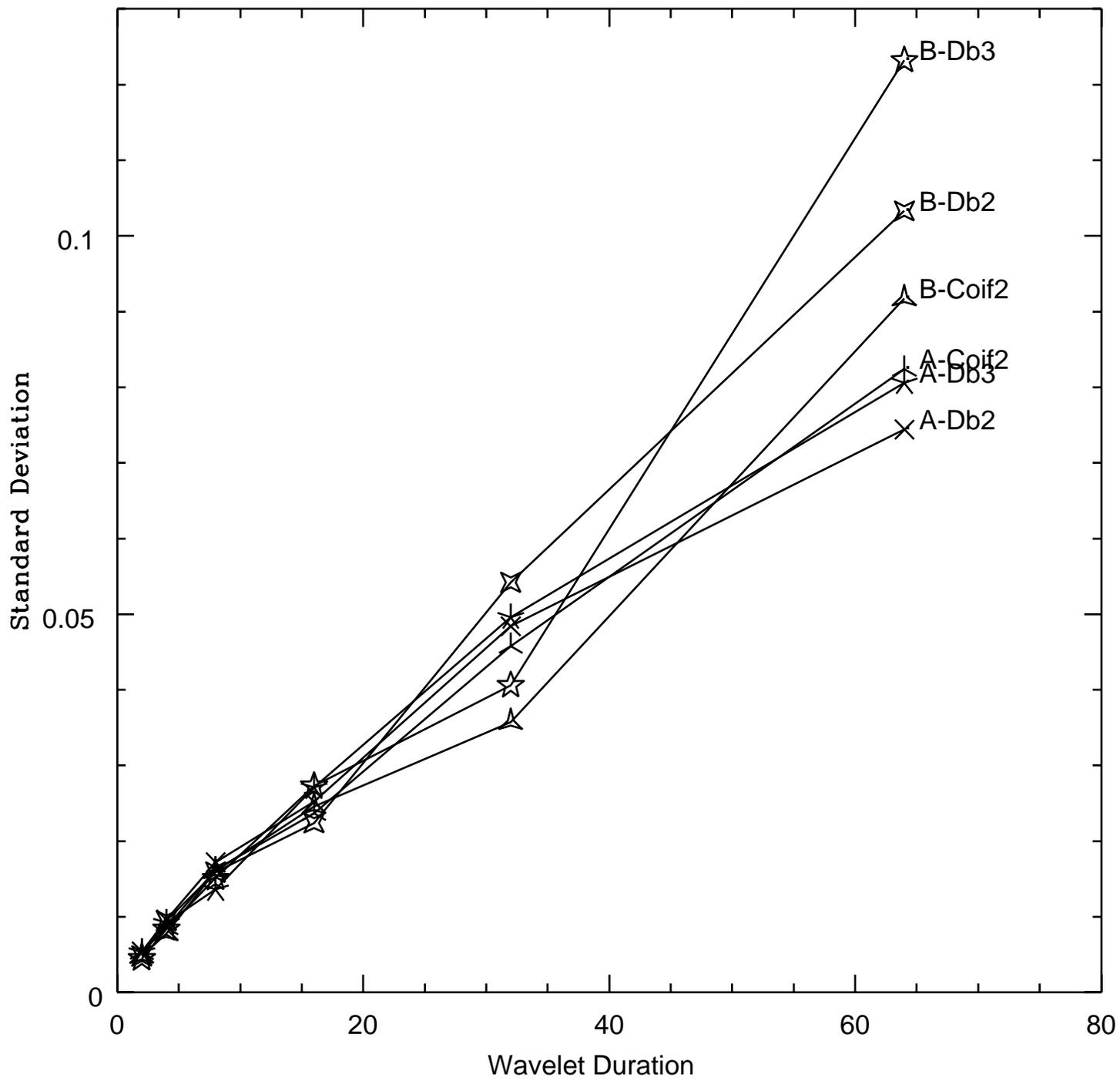
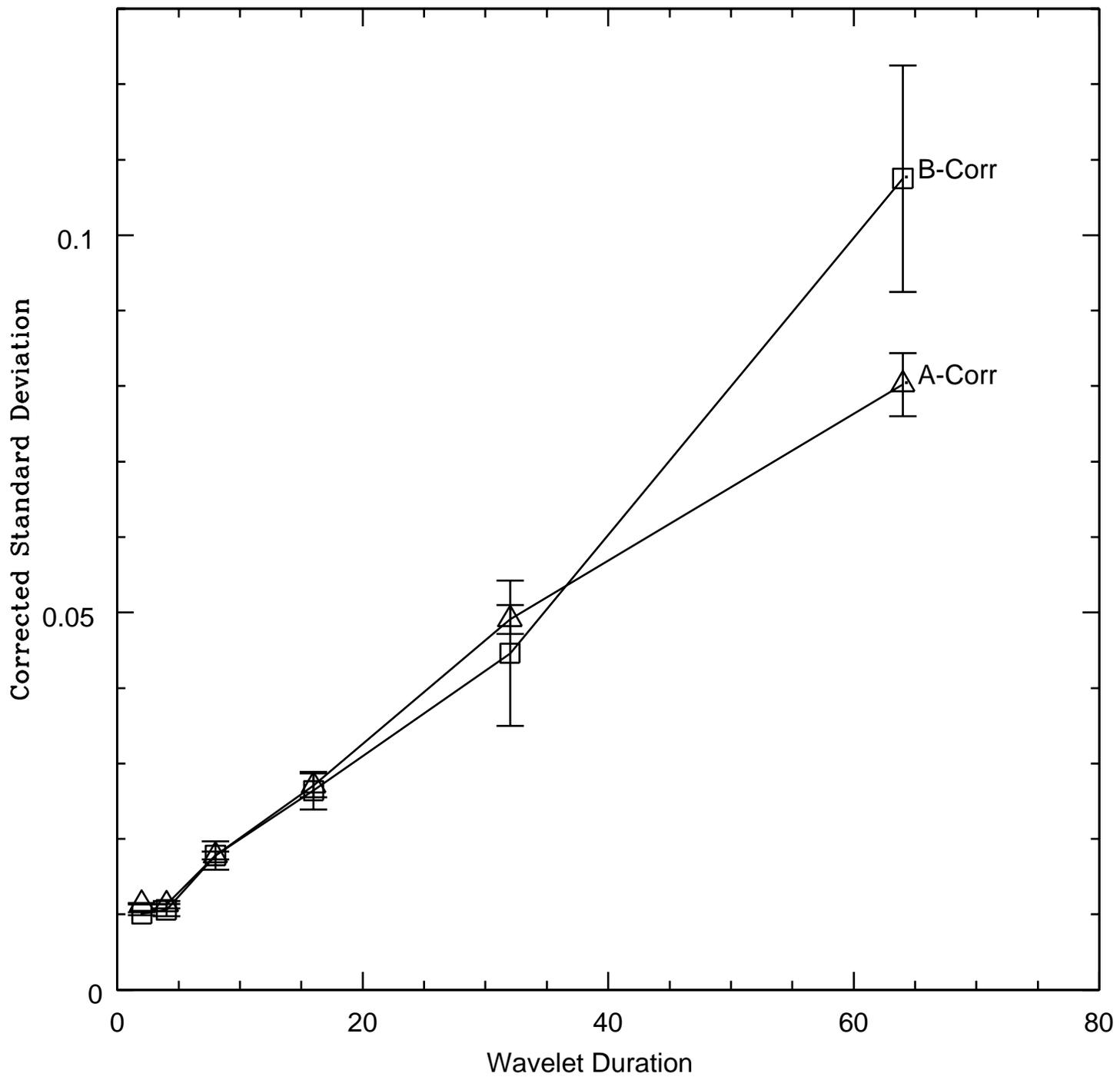
\end{document}